# Multilevel nonvolatile optoelectronic memory based on memristive plasmonic tunnel junctions

*Pan Wang\*, Mazhar E. Nasir, Alexey V. Krasavin, Wayne Dickson and Anatoly V. Zayats\**

Department of Physics and London Centre for Nanotechnology, King's College London, Strand, London WC2R 2LS, UK

\*Correspondence to: pan.wang@kcl.ac.uk, a.zayats@kcl.ac.uk

**Highly efficient information processing in brain is based on processing and memory components called synapses, whose output is dependent on the history of the signals passed through them. Here we have developed an artificial synapse with both electrical and optical memory effects using reactive tunnel junctions based on plasmonic nanorods. In an electronic realization, the electrons tunneled into plasmonic nanorods under low bias voltage are harvested to write information into the tunnel junctions via hot-electron-mediated chemical reactions with the environment. In an optical realization, the information can also be written optically by external light illumination to excite hot electrons in plasmonic nanorods. The stored information is non-volatile and can be read in both realizations either electrically or optically by measuring the resistance or inelastic-tunnelling-induced light emission, respectively. These memristive light-emitting plasmonic tunnel junctions can be used as memory, logic units or artificial synapses in future optoelectronic or neuromorphic information systems.**

Highly efficient information processing in brain utilises multilevel (as opposed to binary used in modern digital computers) logic components called synapses. A memristive device (or memristor) is a resistive electrical element with resistance depending on the history of the applied electrical signals[1,2], and can, therefore, be used as a memory element for the storage of information or as an artificial synapse to emulate biological synapses. Since the first experimental realization based on a metal/oxide/metal (Pt/TiO$_2$/Pt) structure in 2008 [2], memristors have attracted extensive interest due to their important application in next-generation non-volatile memory, signal processing, reconfigurable logic devices, and neuromorphic computing[3–14]. By integrating light-emitting or plasmonic properties with memristive devices, optical memristors have also been demonstrated[15–18]. Most of the memristive devices are based on metal-insulator-metal structures to achieve resistance switching by dynamically configuring the insulating layer (e.g., the formation/annihilation of nanoscale conductive filament), showing low operation voltages (several volts), short set/reset times (<100 ns), and good endurance. The oxide-based insulating layer is typically limited to thicknesses greater than 3 nm[6–9,12–18]. Further reduction of the insulating layer thickness is demanded in order to reduce the device size, operating voltage and energy consumption[3,5]. However, in metal-insulator-metal structures with insulator thickness of less than 2 nm, the occurrence of quantum-mechanical tunnelling effect can cause electron leakage through the insulating layer, which is believed to be a detrimental effect for further downscaling of electronic components such as transistors in integrated circuits.

Here, we demonstrate optoelectronic memristive devices by taking advantage of the electron tunnelling effect. Based on metal-polymer-metal tunnel junctions, we show the simultaneous multistate switching of the resistance and built-in light emission of the junctions, which is realized both electrically and optically by programming the junctions via hot-electron-mediated chemical reactions controlled by the environment. Light-emitting tunnel junction (Fig. 1a), an optoelectronic analog of a biological synapse, was constructed based on a plasmonic nanorod. During the tunnelling process (Fig. 1b), the inelastically tunneled electrons excite plasmons in the nanorod which can subsequently decay radiatively into photons, while those electrons that tunnel elastically, generate hot electrons in the tips of nanorods which can be harvested for the multilevel writing of the junction state. The information stored is non-volatile and can be read both electrically and optically by interrogating the resistance and emission intensity. Optical coding of the tunnel junctions is also possible using the hot-electrons generated in the tunnel junctions by an external illumination. Controlling a gas environment of the tunnel junctions can be used to program the memristor response.

Experimentally, tunnel junctions were constructed in a plasmonic nanorod array (Fig. 1c), which was fabricated by electrodeposition of Au into porous alumina templates (see Supplementary Section 1). Figure 1d presents the cross-sectional view of a plasmonic nanorod array, clearly showing the Au nanorods embedded in the alumina template. The diameter, length, and separation of the nanorods are approximately 65, 480, and 105 nm, respectively. Metal-polymer-metal tunnel junctions were constructed on the surface of the nanorod metamaterial by using a monolayer of poly-L-histidine (PLH) as the tunnel barrier and 'storage' layer (used as a reactant to store information via reconfigurable chemical reactions), and a droplet of eutectic gallium indium (EGaIn) as the top electrode (see Methods and Supplementary Section 2 for details). Each Au nanorod forms a tunnel junction with the top EGaIn contact (Fig. 1a), creating an array of tunnel junctions (Supplementary Section 3) with density determined by the density of Au nanorod array on the order of ~$10^{10}$ cm$^{-2}$, which is close to the density of synapses in human brain (~$10^{14}$ in total). Nonlinear character of current-voltage characteristic (Fig. 1e) confirms the tunnelling of electrons through the metal-polymer-metal junctions[19]. Upon the application of a forward bias, light emission was observed from the substrate side of the device, which is due to the radiative decay of plasmons excited in the nanorod metamaterial (Fig. 1b)[20–25]. The recorded emission spectra (having a linewidth of ~200 nm) as a function of the applied bias are shown in Fig. 1f. With the increase of the bias, the emission intensity increases gradually, accompanied by a blue-shift of the emission peaks following the quantum cut-off law $h\nu_{phot} \leq eV_b$ [20].

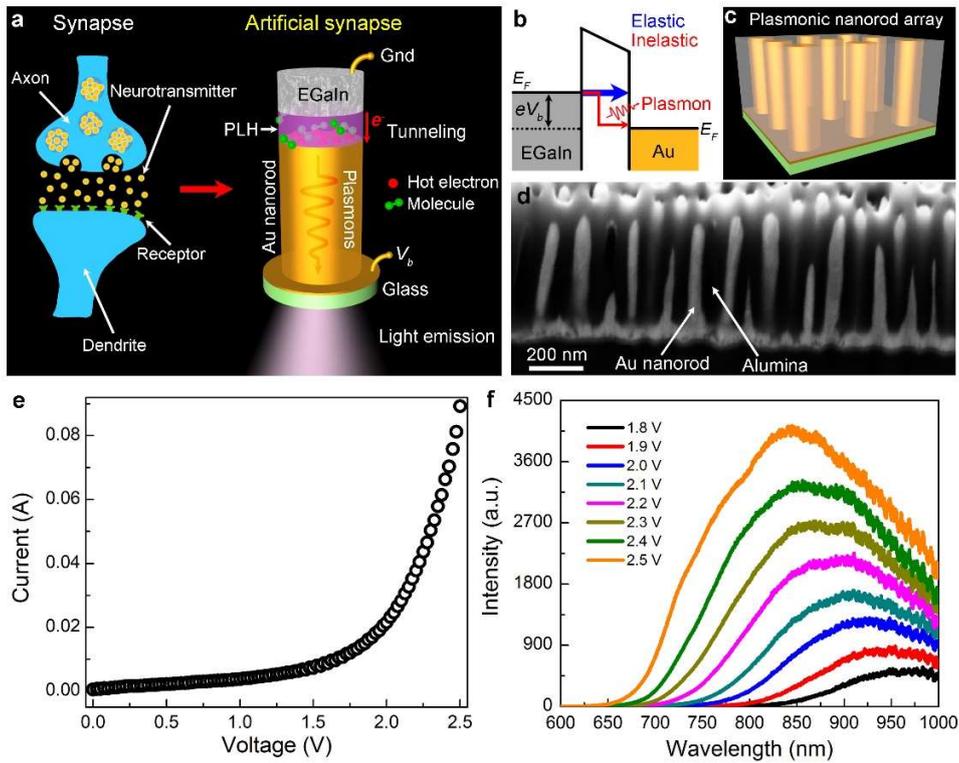

**Figure 1 | Memristor structure and light emission properties. a**, Schematic diagram of the memristive light-emitting tunnel junction (right), which is an optoelectronic analogue to a synapse (left). **b**, An energy level diagram of the tunnel junction with a bias of $V_b$. **c**, Schematic diagram of the plasmonic nanorod array used to realize multiple tunnel junctions. **d**, Cross-sectional SEM view of a nanorod array. **e**, Measured current-voltage characteristic of a tunnelling device fabricated using the array shown in **d**. **f**, Measured emission spectra of the tunnelling device as a function of the applied forward bias.

During the tunnelling process, the majority of electrons (~99%) tunnel elastically (Fig. 1b)[20–25], appearing as hot electrons[26,27] in the tips of Au nanorods, which can be used for programming the state of the tunnel junctions via hot-electron-activated chemical reactions[28,29]. To use the hot-electron effects, the tunnelling device was put into a gas chamber under a bias of 2.5 V, with the tunnelling current and emission spectrum monitored simultaneously. The device was first stabilized in 2% $H_2$ in $N_2$, then, upon switching of a chamber environment to air, the tunnelling current decreased gradually down to two thirds of the original value (Fig. 2a). At the same time, the integrated light emission intensity increased gradually to twice the original value. The changes in the tunnelling current and emission intensity reflect a change in the junction state, which is due to the oxidization of the tunnel junctions by oxygen molecules in air mediated by hot electrons as a PLH monolayer undergoes oxidative dehydrogenation and coupling reactions[25].

The resistance and emission intensity of the tunnelling device depends on the total number of the tunneled electrons (Fig. 2b) since the state of tunnel junctions is dependent on the history of the tunnelling process, particularly on how many electrons have traversed the junctions before, demonstrating the memory effect similar to biological synapses. During the reaction, the device was brought from a low resistance state (~20 Ω) to a high resistance state (~29 Ω), with a simultaneous change in the light emission from a low intensity to high intensity state (~80% increase in intensity). In this case, the written state of the tunnel junctions can be read out both electrically and optically, which is attractive for use as memory devices or artificial synapses, not only in electronic but also in optoelectronic systems. Moreover, compared with the existing optical memristors which require external light sources for the optical readout[16–18,30], the plasmonic tunnel junctions have nanoscale built-in plasmonic light sources, providing advantages for the dramatic reduction in device size and power consumption. Normally, the emission intensity changes linearly with the tunnelling current, however, the emission intensity shows an opposite trend to that of the current during the reaction. This can be understood considering the evolution of the estimated inelastic tunnelling efficiency during the programming process (Fig. 2c, see Supplementary Section 4 for details). During the reaction of the tunnel junctions with oxygen molecules, the inelastic tunnelling efficiency increases gradually, resulting in the increased light emission intensity despite the gradual decrease of the tunnelling current.

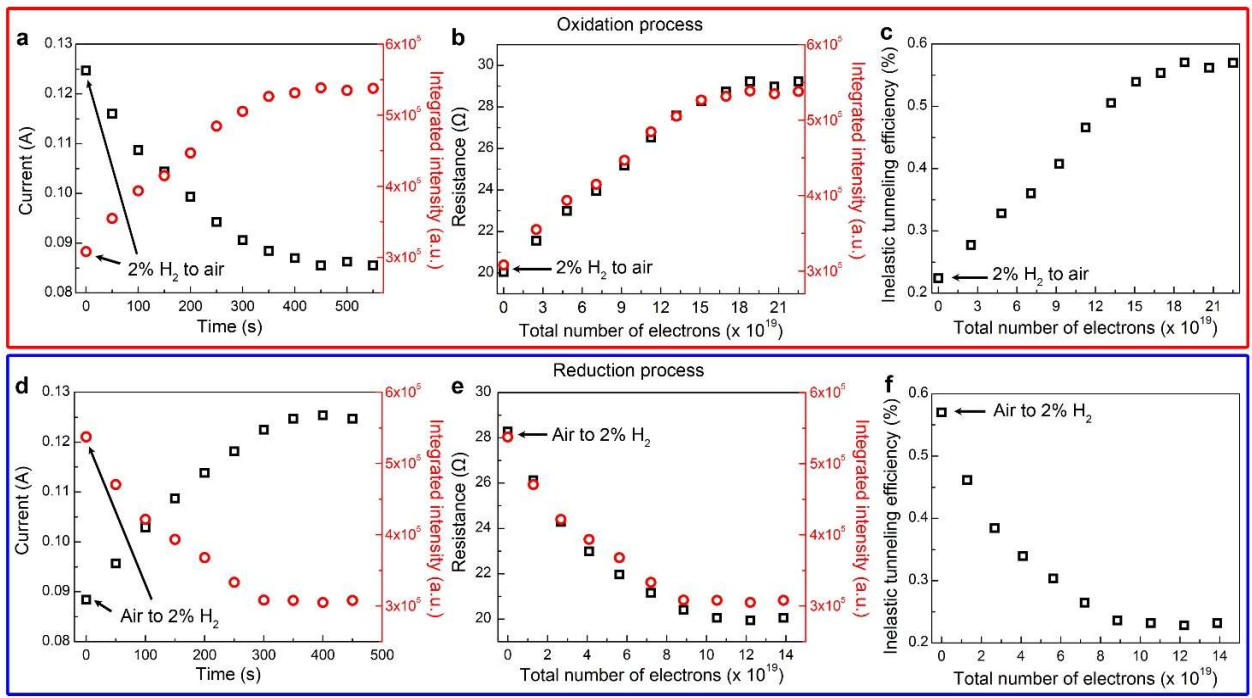

**Figure 2 | Hot-electron-mediated programming of electrical and optical properties. a**, Time-dependent evolution of the tunnelling current and integrated emission intensity during the hot-electron-mediated reaction of tunnel junctions with oxygen (environment switched from 2% $H_2$ gas to air). **b,c**, Dependence of the resistance and integrated emission intensity (**b**), and the inelastic tunnelling efficiency (**c**) on the total number of the tunneled electrons. **d**, Time-dependent evolution of the tunnelling current and integrated emission intensity during the hot-electron-mediated reaction of oxidized tunnel junctions with hydrogen (environment switched from air to 2% $H_2$). **e,f**, Dependence of the resistance

and integrated emission intensity (**e**), and the inelastic tunnelling efficiency (**f**) on the total number of the tunneled electrons. $V_b$ is fixed at 2.5 V in all the measurements.

The tunnelling device can be programmed back to the original status by introducing hydrogen molecules into the cell via the hot-electron-mediated reduction of the oxidized tunnel junctions (Fig. 2d–f). The resistance, integrated emission intensity, and inelastic tunnelling efficiency (Fig. 2e,f) decreased gradually back to the original value with the continuous supply of the hot electrons and hydrogen molecules, highlighting the ability to reversibly programme the tunnelling device. The dynamics of the light-emitting reactive tunnel junctions can be associated with long-term potentiation/depression processes of synapses in a biological neural network. Different from ferroelectric or magnetic tunnel junction based memristors exploiting tunnel electroresistance or magnetoresistance effects[10,11], the light-emitting plasmonic tunnel junctions exploit elastically tunneled electrons for the writing of information and inelastically tunneled electrons for the optical readout, providing programmability of the response and sensitivity to the environment.

As discussed above, the state of the tunnel junctions is highly dependent on the number of the tunneled electrons and the environment. By controlling the supply of hot electrons or reactants (oxygen or hydrogen), the tunnelling device can be latched to different intermediate states. For example, as shown in Fig. 3a, the resistance of the device was switched from ~20 Ω (level L) to 22 (level 1), 26 (level 2), and 29 Ω (level H), respectively, by controllably introducing oxygen molecules into the chamber for the oxidization of the tunnel junctions. When the required state was achieved, pure nitrogen (employed as a nonreactive environment) was introduced into the chamber to remove oxygen molecules to latch the state of the junctions. Under the nonreactive environment of nitrogen, the state of the junctions was maintained when the bias was switched off, showing the non-volatility. Accordingly, the light emission from the device was also latched to different intermediate levels (Fig. 3b). Benefitted from the programming mechanism of the reactive tunnel junctions, the states of the junctions may, in principle, be controlled on single electron or molecule level. Instead of carrying out computations based on binary in digital chips, the artificial synapses based on reactive tunnel junctions work in an analog way like neurons in brain that activate in various way depending on the type and number of ions that flow across a synapse.

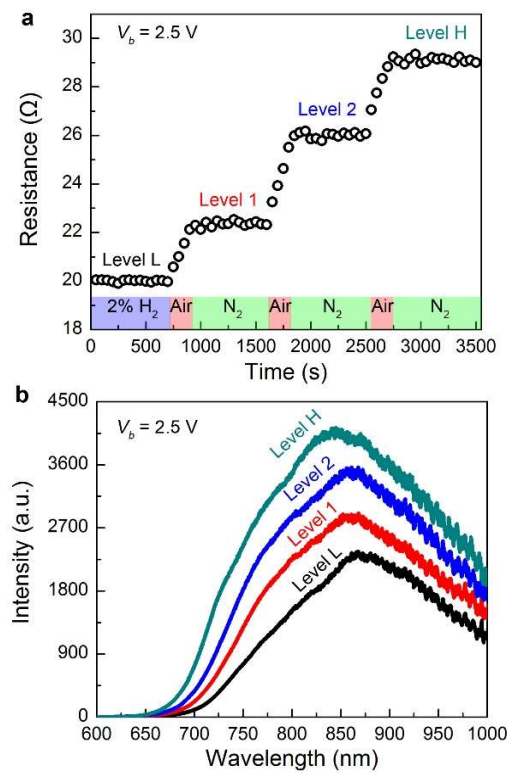

**Figure 3 | Multilevel programming of the electrical and optical properties. a**, Continuous switching of the resistance of the tunnelling device by controlling the environment. The device was first stabilized in 2% $H_2$ ($V_b$ = 2.5 V), then the resistance was switched by the introduction of oxygen molecules and latched by replacing oxygen environment with nitrogen. **b**, Corresponding latched emission spectra of the tunnelling device.

Apart from the electrical programming, the state of the tunnel junctions can be programmed optically. Under external light illumination of the nanorod metamaterial from the substrate side, the plasmonic modes in the metamaterial are excited. Figure 4a shows simulated electric field and current distributions in the unit cell of the metamaterial for the illumination wavelength of 600 nm (see Methods for details). The plasmonic excitation exists across the whole nanorod length and hot electrons are generated in both tips of the Au nanorods, which can be used for the activation of chemical reactions in the tunnel junctions. In order to demonstrate this (Fig. 4b), the device was first stabilized in 2% $H_2$ in $N_2$ under 2.5 V (period 1). The state of the junctions was unchanged when the environment was switched to air under zero bias (period 2) due to the lack of hot electrons for the reaction (under applied bias the gradual rise of the tunnel resistance to level H was observed as expected (period 3)). The state was programmed back to the low resistance level (period 4) by introducing 2% $H_2$ back into the chamber under applied bias. However, when the environment was switched to air under zero bias but the metamaterial was illuminated by a white light (period 5), the state of the junctions was programmed to the high resistance level (confirmed by the stable resistance after the removal of illumination under a bias of 2.5 V (period 6)). The spectra of latched light emission correspond to the level L and optically switched level H' agree well with the emission spectra from the electrically programmed low and high resistance levels (cf., Fig. 4c and Fig. 3b). The ability of optical coding provides an alternative choice for the writing of information with advantages such as wireless and wavelength-dependent control.

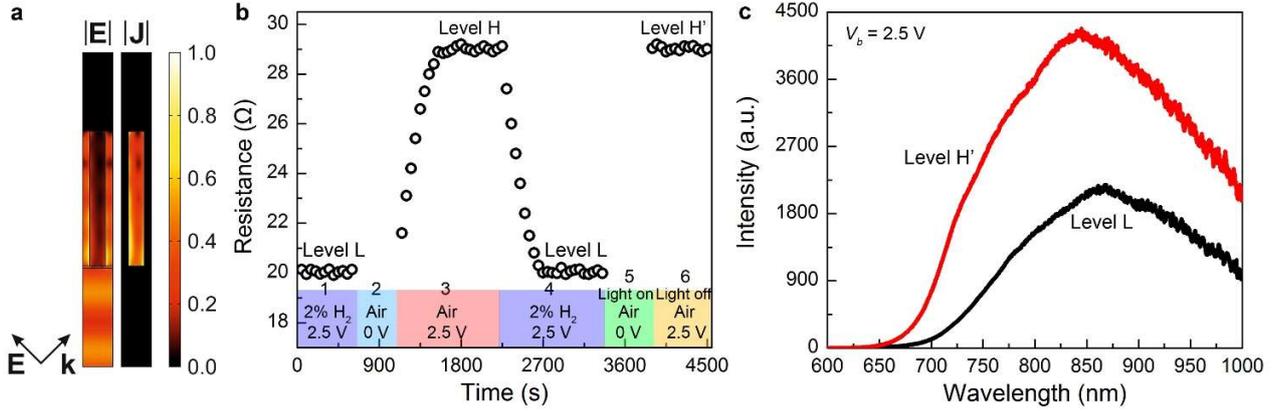

**Figure 4 | Optical programming of the tunnelling device. a**, Simulated electric field |E| and current |J| distributions in the unit cell of the metamaterial for the illumination wavelength of λ = 600 nm. **b**, Evolution of the resistance of the tunnelling device under different conditions as indicated at the bottom. White light illumination in the spectral range 500-750 nm and power density ~0.03 W cm$^{-2}$ was used during period 5. **c**, Emission spectra of the tunnelling device measured at the resistance level L and the optically programmed resistance level H'.

In conclusion, we have investigated the electrical and optical memory effects in reactive tunnel junctions. The high density of tunnel junctions and scalability provided by the plasmonic nanorod array make the proposed approach an attractive platform for the construction of 'brain on a chip' and neuromorphic computing devices. Flexibility of the approach can be further exploited using new chemical reactions and other switching mechanisms, such as formation/annihilation of conductive filament, to further control the operation. The reactive plasmonic tunnel junctions can be integrated directly with plasmonic or silicon waveguides for the application as memory, logic units, and artificial synapses in optoelectronic systems, scaled down to single junctions if required. The light-emitting reactive tunnel junctions have the potential to become important building blocks of memories, logic units, or artificial synapses in optoelectronic or neuromorphic computing systems.

## Acknowledgement

This work has been funded in part by the Engineering and Physical Sciences Research Council (UK) and the European Research Council iPLASMM project (321268). A.V.Z. acknowledges support from the Royal Society and the Wolfson Foundation.

## Author contributions

A.V.Z. and P.W. conceived the study. P.W. constructed the experiment, performed the measurement and analysed the data. M.E.N. and W.D. fabricated the nanorod metamaterials. A.V.K. performed numerical simulations. All the authors discussed the results and co-wrote the paper.

## Additional information

Correspondence and requests for materials should be addressed to A.V.Z and P.W.

## Competing financial interests

The authors declare no competing financial interests.

# References

1. Chua L. O. Memristor-the missing circuit element. *IEEE Trans. Circuit Theory* **18**, 507–519 (1971).

2. Strukov D. B., Snider G. S., Stewart D. R. & Williams R. S. The missing memristor found. *Nature* **453**, 80–83 (2008).

3. Yang J. J., Strukov D. B. & Stewart D. R. Memristive devices for computing. *Nat. Nanotechnol.* **8**, 13–24 (2013).

4. Parkin, S. & Yang, S. -H. Memory on the racetrack. *Nat. Nanotechnol.* **10**, 195–198 (2015).

5. Zidan M. A., Strachan J. P. & Lu W. D. The future of electronics based on memristive systems. *Nat. Electron.* **1**, 22–29 (2018).

6. Yang J. J. *et al*. Memristive switching mechanism for metal/oxide/metal nanodevices. *Nat. Nanotechnol.* **3**, 429–433 (2008).

7. Li C. *et al*. Analogue signal and image processing with large memristor crossbars. *Nat. Electron.* **1**, 52–59 (2018).

8. Borghetti J. *et al*. 'Memristive' switches enable 'stateful' logic operations via material implication. *Nature* **464**, 873–876 (2010).

9. Jo S. H. *et al*. Nanoscale memristor device as synapse in neuromorphic systems. *Nano Lett.* **10**, 1297–1301 (2010).

10. Krzysteczko P., Münchenberger J., Schäfers M., Reiss G., Thomas A. The memristive magnetic tunnel junction as a nanoscopic synapse-neuron system. *Adv. Mater.* **24**, 762–766 (2012).

11. Kim D. J. *et al*. Ferroelectric tunnel memristor. *Nano Lett.* **12**, 5697–5702 (2012).

12. Wu C. X., Kim T. W., Choi H. Y., Strukov D. B. & Yang J. J. Flexible three-dimensional artificial synapse networks with correlated learning and trainable memory capability. *Nat. Commun.* **8,** 752 (2017).

13. Schneider M. L. *et al*. Ultralow power artificial synapses using nanotextured magnetic Josephson junctions. *Sci. Adv.* **4,** e1701329 (2018).

14. Choi S. *et al*. SiGe epitaxial memory for neuromorphic computing with reproducible high performance based on engineered dislocations. *Nat. Mater.* **17,** 335–340 (2018).

15. He C. L. *et al*. Tunable electroluminescence in planar graphene/$SiO_2$ memristors. *Adv. Mater.* **25,** 5593–5598 (2013).

16. Emboras A. *et al*. Nanoscale plasmonic memristor with optical readout functionality. *Nano Lett.* **13,** 6151–6155 (2013).

17. Emboras A. *et al*. Atomic scale plasmonic switch. *Nano Lett.* **16,** 709–714 (2016).

18. Cheng Z. G., Ríos C., Pernice W. H. P., Wright C. D. & Bhaskaran H. On-chip photonics synapse. *Sci. Adv.* **3,** e1700160 (2017).

19. Simmons J. G. Generalized formula for the electric tunnel effect between similar electrodes separated by a thin insulating film. *J. Appl. Phys.* **34,** 1793–1803 (1963).

20. Lambe J. & McCarthy S. L. Light emission from inelastic electron tunneling. *Phys. Rev. Lett.* **37,** 923–925 (1976).

21. Kern J. *et al*. Electrically driven optical antennas. *Nat. Photon.* **9,** 582–586 (2015).

22. Parzefall M. *et al*. Antenna-coupled photon emission from hexagonal boron nitride tunnel junctions. *Nat. Nanotechnol.* **10,** 1058–1063 (2015).

23. Du W. *et al*. On-chip molecular electronic plasmon sources based on self-assembled monolayer. *Nat. Photon.* **10,** 274–280 (2016).

24. Du W., Wang T., Chu H. –S. & Nijhuis C. A. Highly efficient on-chip direct electronic-plasmonic transducers. *Nat. Photon.* **11,** 623–627 (2017).

25. Wang P., Krasavin A. V., Nasir M. E., Dickson W. & Zayats A. V. Reactive tunnel junctions in electrically driven plasmonic nanorod metamaterials. *Nat. Nanotechnol.* **13,** 159–164 (2018).

26. Brongersma M. L., Halas N. J. & Nordlander P. Plasmon-induced hot carrier science and technology. *Nat. Nanotech.* **10,** 25–34 (2015).

27. Harutyunyan H. *et al*. Anomalous ultrafast dynamics of hot plasmonic electrons in nanostructures with hot spots. *Nat. Nanotechnol.* **10,** 770–774 (2015).

28. Mubeen S. *et al*. An autonomous photosynthetic device in which all charge carriers derive from surface plasmons. *Nat. Nanotechnol.* **8,** 247–251 (2013).

29. Zhai Y. M. *et al*. Polyvinylpyrrolidone-induced anisotropic growth of gold nanoprisms in plasmon-driven synthesis. *Nat. Mater.* **15,** 889–895 (2016).

30. Hoessbacher C. *et al*. The plasmonic memristor: a latching optical switch. *Optica* **1,** 198–202 (2014).

## Methods

**Fabrication.** Plasmonic nanorod metamaterials were fabricated by electrodeposition of Au into substrate-supported porous alumina templates[25,31]. Metamaterial-based light emitting tunnel junctions were fabricated as follows: firstly, a nanorod metamaterial was chemically etched in a 3.5% $H_3PO_4$ solution at 35 °C to make the surrounding $Al_2O_3$ matrix slightly lower than the nanorod tips; secondly, the metamaterial with exposed nanorod tips were functionalized with a monolayer of PLH ($M_w$ = 5,000—25,000, Sigma-Aldrich) via self-assembly; finally, a droplet of EGaIn (≥ 99.99% trace metals basis, Sigma-Aldrich) was added onto the surface of the metamaterial to form an array of metal-PLH-metal tunnel junctions.

**Numerical simulations.** Numerical simulations of the near-field distributions of the electric field inside the nanorod metamaterial and the associated electric current in the nanorods were performed using a finite element method (Comsol Multiphysics software). The metamaterial was illuminated from the substrate side by a plane wave at an angle of incidence of 45°. The distribution of the nanorods in the metamaterial was approximated with a square array, which allowed to simulate the entire system modeling a unite cell with properly defined Floquet boundary conditions on the unit cell sides, determined by a phase delay acquired by the incident plane wave while travelling between the corresponding faces. To ensure the absence of back-reflection, a perfectly matched layer was implemented on the illumination side. At the opposite (EGaIn) side, this was not needed due to metallic nature of the latter. Experimentally measured data with a mean free path correction of 3 nm, related to the properties of electrochemically derived Au, were used for gold[32], experimental tabulated data was also used for $Al_2O_3$ matrix[33], $SiO_2$ substrate[34] and $Ta_2O_5$ adhesion layer[35]. Optical properties of EGaIn were approximated by the Drude model, while refractive indices of gallium oxide, naturally appearing on the EGaIn surface, and PLH polymer were taken as non-dispersive in the studied wavelength range and equal to 1.89 and 1.565, respectively.

**References**

31. Evans P. *et al*. Growth and properties of gold and nickel nanorods in thin film alumina. *Nanotechnology* **17,** 5746–5753 (2006).

32. Johnson P. B. & Christy R. W. Optical constants of the noble metals. *Phys. Rev. B* **6,** 4370–4379 (1972).

33. Malitson I. H. & Dodge M. J. Refractive index and birefringence of synthetic sapphire. *J. Opt. Soc. Am.* **62,** 1405 (1972).

34. Malitson I. H. Interspecimen comparison of the refractive index of fused silica. *J. Opt. Soc. Am.* **55,** 1205-1209 (1965).

35. Bright T. J. *et al*. Infrared optical properties of amorphous and nanocrystalline $Ta_2O_5$ thin films. *J. Appl. Phys.* **114,** 083515 (2013).

# Supplementary Information

# Multilevel nonvolatile optoelectronic memory based on memristive plasmonic tunnel junctions

*Pan Wang\*, Mazhar E. Nasir, Alexey V. Krasavin, Wayne Dickson and Anatoly V. Zayats\**

Department of Physics and London Centre for Nanotechnology, King's College London, Strand, London WC2R 2LS, UK

Correspondence to: pan.wang@kcl.ac.uk, a.zayats@kcl.ac.uk

## S1. Fabrication of plasmonic nanorod metamaterials

The plasmonic nanorod metamaterials were fabricated by electrodeposition of Au into substrate-supported porous alumina templates [31]. The substrate is a multilayered structure comprised of a glass slide (1 mm in thickness), a tantalum oxide adhesive layer (10 nm in thickness), and an Au film (7 nm in thickness) acting as a working electrode for the electrochemical reaction. An aluminum film (~500 nm in thickness) is then deposited onto the substrate by planar magnetron sputtering, which is subsequently anodized in oxalic acid (0.3 M) at 40 V to produce the porous alumina template by two-step anodization. The diameter, separation and ordering of the Au nanorods in the assembly are controlled by the conditions of anodization. The electrodeposition of Au is performed with a three-electrode system using a non-cyanide solution. The length of Au nanorods is controlled by the electrodeposition time. In this work, the length of Au nanorods is slightly shorter than the height of alumina template. The metamaterials were then washed several times in deionized water (DI water) and stored in 200 proof ethanol for future use.

## S2. Fabrication of metal-polymer-metal tunnel junctions

Figure S1 shows the schematic diagram of the fabrication of metal-polymer-metal tunnel junctions based on a plasmonic nanorod metamaterial. In the first step, a wet chemical etching method was used to remove some of the alumina matrix to make the tips of Au nanorods slightly higher than the surrounding alumina. Briefly, the nanorod metamaterial stored in ethanol was firstly dried under $N_2$ and then put into an aqueous solution of $H_3PO_4$ (3.5 %) at 35 °C to start the etching. The etching depth can be precisely controlled by the etching time. After the chemical etching, the metamaterial was washed several times in DI water.

In the second step, the exposed Au nanorod tips were functionalized with a monolayer of poly-L-histidine (PLH, Mw 5,000-25,000, Sigma-Aldrich). Briefly, the etched nanorod metamaterial was submerged into a PLH solution (1 mg/mL, pH ~5–6) and incubated for 0.5 h. Due to the high affinity of PLH to Au surface and the positive charging of protonated PLH in solution, a monolayer of PLH self-assembled onto the exposed Au nanorod tips. The metamaterial was then washed several times in DI water to remove weakly bound PLH and dried under $N_2$.

Finally, a droplet of EGaIn was added onto the surface of the nanorod metamaterial to form an array of metal-PLH-metal tunnel junctions, and then two Au wires were connected to the bottom Au film and the EGaIn droplet for the application of bias across the tunnel junctions.

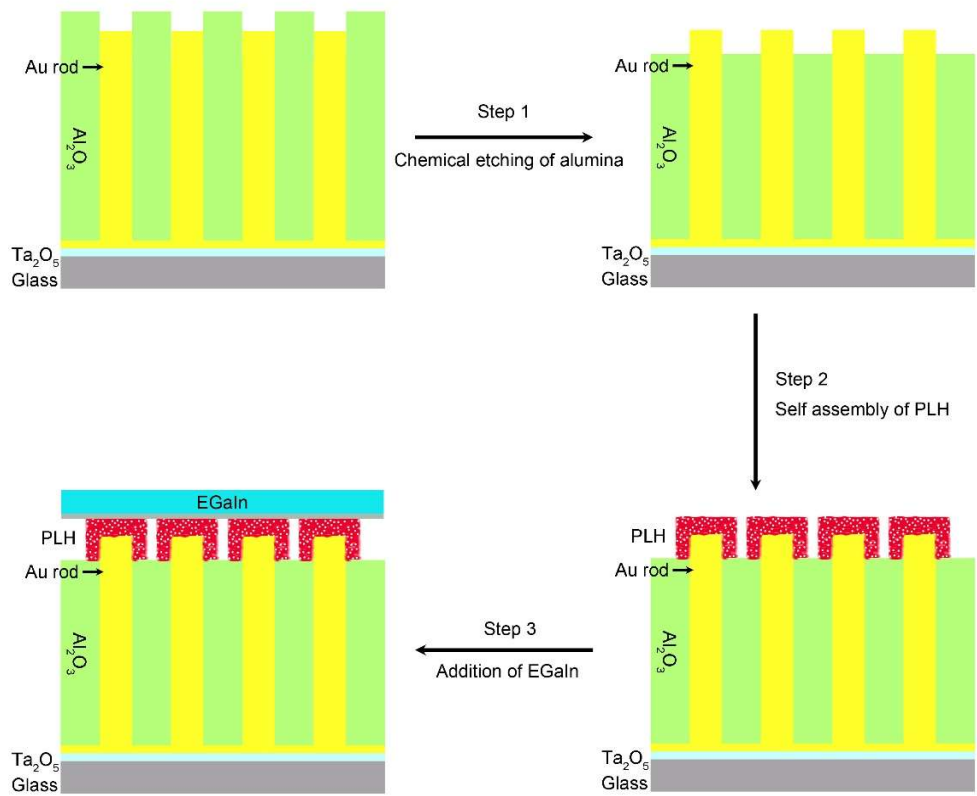

**Figure S1**. Schematic diagram showing the steps for the fabrication of metal-polymer-metal tunnel junctions.

## S3. Nanorod-metamaterial-based tunnelling device

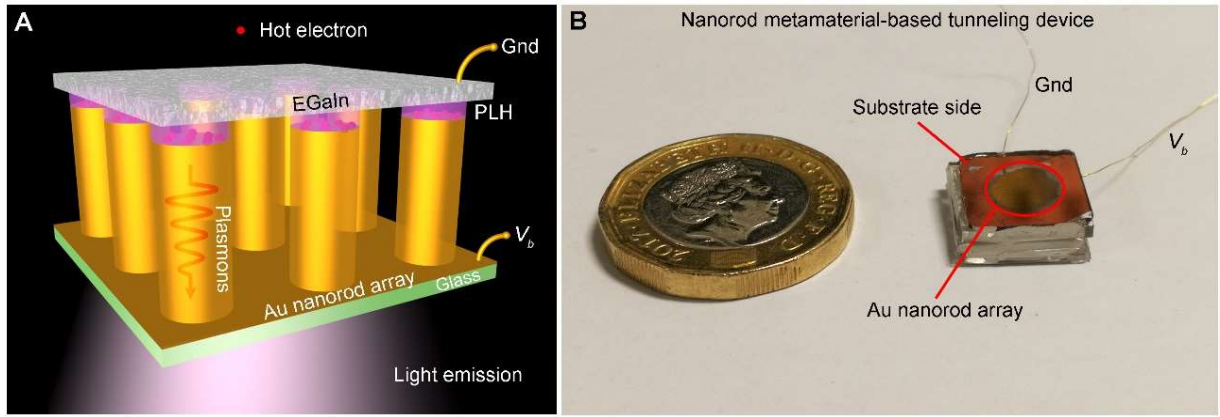

**Figure S2 | Nanorod-metamaterial-based tunnelling device. a**, Schematic diagram of tunnelling junction array constructed on a Au nanorod metamaterial. **b**, Photograph of a nanorod-metamaterial-based tunnel device.

## S4. Estimation of inelastic tunnelling efficiency

In the nanorod-metamaterial-based tunnel junctions, the light emission is due to the radiative decay of plasmonic modes excited by the inelastic tunnelling electrons. In this case, the relation between inelastic tunnelling efficiency ($\eta_{inel} = \Gamma_{inel}/\Gamma_{tot}$, where $\Gamma_{inel}$ and $\Gamma_{tot}$ are inelastic and total tunnelling rates, respectively) and electron-to-photon conversion efficiency ($\eta_{el-ph}$) can be written as: $\eta_{el-ph} = \eta_{inel} \cdot \eta_{ant}$, where $\eta_{ant}$ is the antenna radiation efficiency ($\eta_{ant} = P_{rad}/P_{tot}$, defining how much power from the excited plasmonic modes is radiated in light). The electron-to-photon conversion efficiency can be estimated from the ratio of emitted photons (evaluated from the measured emission power, assuming all the emitted photons have the same wavelength of 850 nm) to injected electrons (evaluated from the measured tunnelling current under 2.5 V forward bias) during each period of measurement. The antenna radiation efficiency $\eta_{ant}$ can be evaluated from the numerical simulations to be ~3.5 × 10$^{-4}$ at 850 nm.

We can plot the evolution of inelastic tunnelling efficiency during the memristor programming process using the formula: $\eta_{inel} = \eta_{el-phot}/\eta_{ant}$.